%% file: main.tex
\documentclass[%
 preprint,
superscriptaddress,
showpacs,
showkeys,
 amsmath,amssymb,
 aps,
prb,
]{revtex4-1}
\usepackage{graphicx,amssymb,amsmath,epsfig}
\usepackage{color}
\usepackage{comment}

\usepackage{hyperref}
\hypersetup{
    colorlinks=true,
    linkcolor=blue,
    filecolor=blue,      
    urlcolor=blue,
    citecolor=blue,
    pdftitle={Overleaf Example},
    pdfpagemode=FullScreen,
    }

\begin{document}

\title{2D Porphyrazine: A New Nanoporous Material}
\author{R. M. Tromer}
\affiliation{Applied Physics Department, University of Campinas, Campinas, S\~ao Paulo, Brazil.}
\affiliation{Center for Computing in Engineering and Sciences, University of Campinas, Campinas, S\~ao Paulo, Brazil.}
\author{M. L. Pereira Junior}
\affiliation{Department of Electrical Engineering, University of Bras\'{i}lia 70919-970, Brazil.}
\author{L. A. Ribeiro Junior}
\affiliation{Institute of Physics, University of Bras\'ilia, 70910-900, Bras\'ilia, Brazil.}
\author{D. S. Galv\~ao}
\affiliation{Applied Physics Department, University of Campinas, Campinas, S\~ao Paulo, Brazil.}
\affiliation{Center for Computing in Engineering and Sciences, University of Campinas, Campinas, S\~ao Paulo, Brazil.}
\date{\today}

\begin{abstract}
Crystalline microporous materials are solids formed by interconnected pores of less than 2 nm in size. Typically, they possess large surface areas desirable for versatile applications such as catalysis, gas adsorption, and energy storage. In the present work, we propose a new porphyrin-based 2D nanoporous crystal, named 2D Porphyrazine (2DP), which is formed by topological assembling H$_{5}$C$_{13}$N$_{4}$ porphyrins. We have considered its monolayer, bi-layer, and molecular crystal (bulk) arrangements. We carried out DFT calculations to investigate 2DP structural and electronic properties. Results show that 2DP is a very stable structure with a direct bandgap of 0.65 eV and significant optical absorption in the visible range. 2DP exhibited satisfactory affinity to lithium atoms. Simulations also showed the existence of proton transfer between nitrogen atoms. It is the first report on the site-specific hydrogen exchange process in 2D crystals. 
\end{abstract}



\maketitle

\section{Introduction}

Carbon-based 2D materials, which include graphene as the most important one \cite{geim2009graphene,novoselov2004electric} and other allotropes \cite{enyashin2011graphene,singh2011graphene}, have been widely investigated to propose new solutions for organic electronics \cite{gupta2011luminscent,han2012extremely,schwierz2010graphene}. Some of the graphene properties, such as high thermal and electrical conductivity and mechanical strength stand out \cite{bonaccorso2010graphene,sensale2015graphene}. However, its null band gap poses some limitations to nanoelectronics since it prevents, for instance, the device from working in the on/off current ratio \cite{liu2013graphene,ning2020flexible}. 

To overcome this problem, new strategies for opening the graphene bandgap \cite{han2007energy,xia2010graphene,cadelano2010elastic} and for designing new 2D materials with a semiconducting bandgap in their pristine form have been investigated \cite{zhang2019art,chhowalla2013chemistry,abrahams1994assembly}. Among these new materials, 2D porphyrin-based crystals are attractive due to their small semiconducting band gaps and nanoporous structure with large area surface, which are crucial traits for versatile applications, such as catalysis \cite{hao2019facile,zhu2021porphyrin}, selectivity \cite{zhang2021historical}, gas adsorption \cite{sharma2019mn}, and energy storage \cite{liao20162d}.          

Recently, 2D porphyrin-based nanoporous materials have been both experimentally \cite{chen2018local,abel2011single,yang2015microporous,zhang2015permanently,xu2020semiconductive,bhunia2017electrochemical,titi2016porphyrin} and theoretically \cite{tromer2021thiophene,chilukuri2020structure,roy2012catalysis,luo2017two,diboron,hamad2015electronic,singh2015theoretical} investigated. Abel and coworkers reported the synthesis of a well-ordered organometallic sheet consisting of 2D polymeric phthalocyanine \cite{abel2011single}. The growth demonstrated on a metal surface can be extended onto a thin insulating film. Phthalocyanine is one of the possible structures that can be formed by porphyrin molecules \cite{goldberg2005crystal}. Other 2D porphyrin-based metal-organic frameworks were also successfully synthesized \cite{lu2017synthesis,chen2021rational,zhou2021enhancing,zhao2018synthesis}. As a general trend, the fabricated materials exhibited good electrocatalytic activity and selectivity. 

Density functional theory (DFT) calculations were also carried out to design new 2D porphyrin-based crystals \cite{tromer2021thiophene,chilukuri2020structure,roy2012catalysis,luo2017two,diboron,hamad2015electronic,singh2015theoretical}. A common feature that can be inferred from these numerical studies, although they are very few in the literature so far, is that 2D porphyrin-based crystals can present semiconducting electronic bandgaps, ranging between 0.6-2.0 eV \cite{chilukuri2020structure,diboron,chen2018local}. These bandgap values are in the range required for materials that can be useful in flexible nanoelectronics. It is, therefore, relevant to propose other 2D porphyrin-based crystals to expand the understanding of their possible use as active layers in these applications.   

In this work, we propose a new porphyrin-based 2D nanoporous crystal (see Figure \ref{fig:structure}), named 2D Porphyrazine (2DP), which is formed by topological assembling H$_5$C$_{13}$N$_{4}$ porphyrins. We use DFT simulations to investigate its structural and electronic properties. 2DP is a direct bandgap semiconductor with an electronic bandgap value of 0.65 eV and exhibits high thermal stability for temperatures up to 2000 K. The absence of negative frequencies in the phonon spectrum confirms its structural stability. The optical absorption of 2DP is intense within the visible region. We also investigated the lithium interaction with 2DP sheets. \textit{ab initio} molecular dynamics (AIMD) simulations revealed that 2DP tends to adsorb these atoms, suggesting good potential for energy storage applications. 

\section{Methodology}

We performed DFT calculations to investigate 2DP electronic and optical properties using the SIESTA code \cite{Soler2002}. We have considered single and bi-layer, as also the molecular crystal (bulk). The simulations were carried out within the generalized gradient approximation (GGA), with the Perdew-Burke-Ernzenhof (PBE) \cite{perdew1996generalized,ernzerhof1999assessment}, as the exchange-correlation functional. van der Waals (vdw-DFT) corrections to describe the exchange-correlation term \cite{vdw1,vdw2,vdw3} was also used. Norm-conserving Troullier-Martins pseudopotential describes the core electrons \cite{Troullier1991}. We used two basis sets, double-zeta plus polarization (DZP) and single-zeta (SZ). Sz was used in the AIMD simulations because they have a higher computational cost.

We used a kinetic energy cut-off of $300$ eV. The k-grid is $6\times6\times1$ for geometry optimization and $30\times30\times1$ for electronic and optical calculations. To prevent spurious interactions among the 2DP layers, we used a vacuum region of $20$ \r{A}. Lattice vectors and atom positions were fully-relaxed in the optimization process. The criteria convergence was of maximum force at each atom less than $1.0\times10^{-3}$ eV/\r{A} and for total energy was $1.0\times10^{-6}$ eV. To investigate the structural stability of the 2DP structure, we performed phonon dispersion calculations based on the force constant algorithm for T${}=0$ K \cite{FC}. 

To obtain the optical properties, we considered an external electric field of 1.0 V/\r{A} along the x and y-directions. This value is characteristic for 2D systems similar to 2DP, such as nitrogenated holey graphene  \cite{Tromer2017}. In this sense, we derived the optical quantities directly from the complex dielectric constant $\epsilon=\epsilon_1+i\epsilon_2$, where $\epsilon_1$ and $\epsilon_2$ are the real and imaginary parts of dielectric constant, respectively.  

From Fermi's golden rule, we obtained the imaginary part of the dielectric constant from the interband optical transitions between valence (VB) and conduction bands (CB),

\begin{equation}
\epsilon_2(\omega)=\frac{4\pi^2}{V_\Omega\omega^2}\displaystyle\sum_{i\in \mathrm{VB}, \, j\in \mathrm{CB}}\displaystyle\sum_{k} W_k \, |\rho_{ij}|^2 \, \delta	(\varepsilon_{kj}-\varepsilon_{ki}- \hbar \omega),
\end{equation}
where $\omega$ is the photon frequency, $V_\Omega$ the unit cell volume, $W_k$ the k-point weight in the reciprocal space, and $\rho_{ij}$ the dipole transition matrix element \cite{tromer2021thiophene}. $\epsilon_1$ and $\epsilon_2$ are related through the Kramers-Kronig relation. The real part of the dielectric constant is expressed as:

\begin{equation}
\epsilon_1(\omega)=1+\frac{1}{\pi}P\displaystyle\int_{0}^{\infty}d\omega'\frac{\omega'\epsilon_2(\omega')}{\omega'^2-\omega^2},
\end{equation}
where $P$ is the principal value of the logarithm. Once the real and imaginary parts of the dielectric constant are obtained, the other relevant optical coefficients, such as the absorption coefficient $\alpha$, the refractive index ($\eta$), and reflectivity ($R$), can be derived by:  

\begin{equation}
\alpha (\omega )=\sqrt{2}\omega\bigg[(\epsilon_1^2(\omega)+\epsilon_2^2(\omega))^{1/2}-\epsilon_1(\omega)\bigg ]^{1/2},
\end{equation}
\begin{equation}
\eta(\omega)= \frac{1}{\sqrt{2}} \bigg [(\epsilon_1^2(\omega)+\epsilon_2^2(\omega))^{1/2}+\epsilon_1(\omega)\bigg ]^{2},
\end{equation}
and
\begin{equation}
R(\omega)=\bigg [\frac{(\epsilon_1(\omega)+i\epsilon_2(\omega))^{1/2}-1}{(\epsilon_1(\omega)+i\epsilon_2(\omega))^{1/2}+1}\bigg ]^2.
\end{equation}

\section{Results}

\subsection{2DP Stability}

\begin{figure}[!htb]
    \centering
    \includegraphics[scale=0.28]{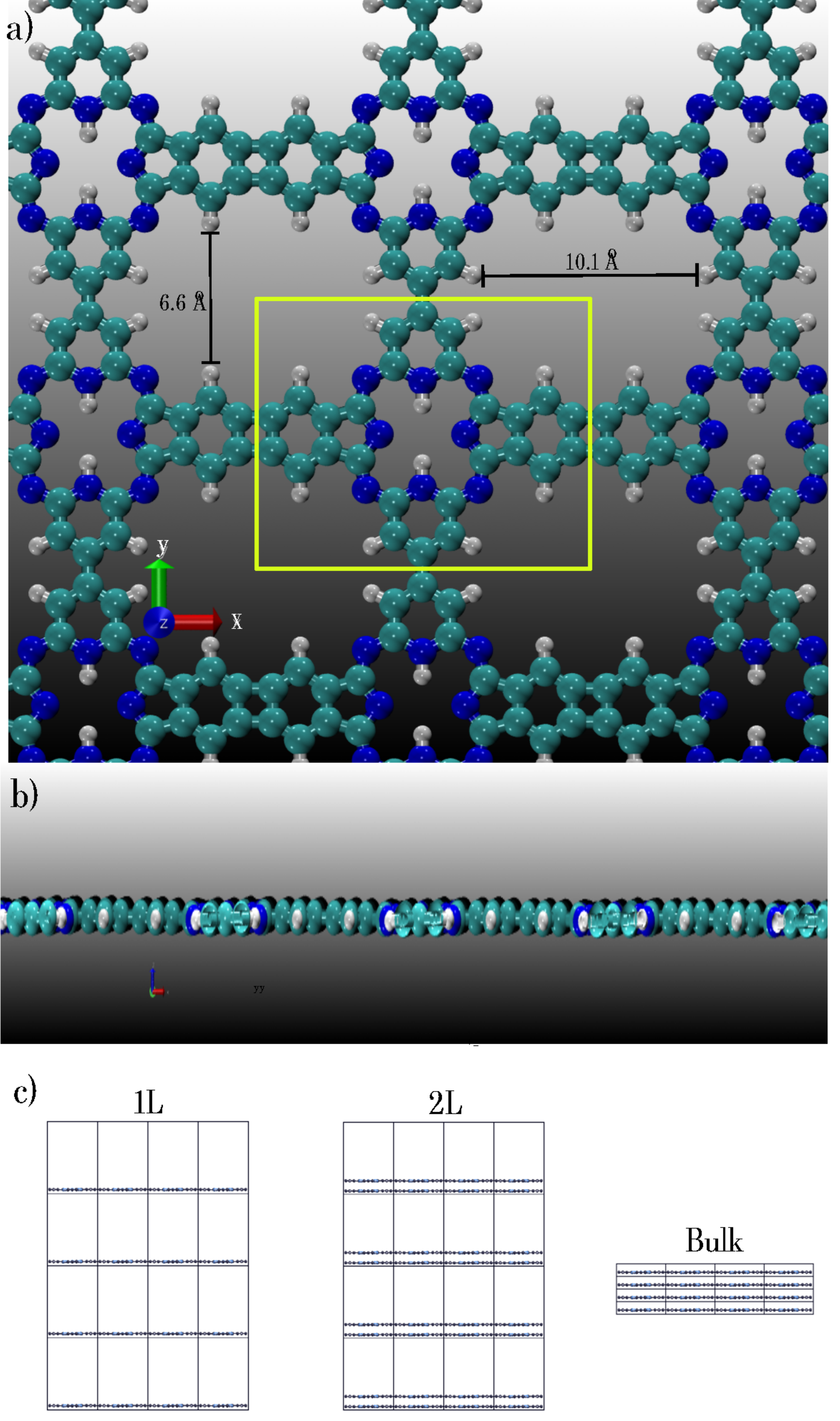}
    \caption{Schematic representation for the optimized 2DP structure. Its top and side views are presented in (a) and (b), respectively. In the color scheme, the carbon, hydrogen, and nitrogen atoms are shown in cyan, white, and blue, respectively. The yellow region highlights the lattice unit cell, and the red and green arrows illustrate the pore size dimensions along the different directions. Panel (c) shows the supercell configuration for the three 2DP models: (1L) single-layer, (2L) bi-layer, and (Bulk) the molecular crystal. For the 1L and 2L simulations, a vacuum buffer layer of 20 \r{A} was used.}
    \label{fig:structure}
\end{figure}

Simulations performed here start from designing the 2DP structure presented in Figure \ref{fig:structure}. The orthorhombic unit cell is highlighted in yellow. It was generated by replacing some C-H with C-C covalent bonds of the hemiporphyrazine molecule previously synthesized \cite{hemiporphyrazine}. The 2DP single-layer shown in Figure \ref{fig:structure}, with composition H$_{5}$C$_{13}$N$_{4}$, has similar optimized parameters when calculated with both GGA and GGA/vdW approximations, as expected: $a=14.53$ \r{A}, $b=11.81$ \r{A}, $c=20.00$ \r{A} for GGA and $a=14.55$ \r{A}, $b=11.80$ \r{A}, $c=20.00$ for vdW, with $\alpha=\beta=\gamma=90^\circ$ for both cases. The average bonds are approximately $R_{C-C}=1.44$ \r{A}, $R_{C-N}=1.35$ \r{A}, $R_{N-H}=1.04$ \r{A} and $R_{C-H}=1.10$ \r{A}. The distance between atoms within the pore can vary from $6.6$ \r{A} to $10.1$ \r{A}. 

The 2DP formation enthalpy values are $-7.7$ and $-8.7$ eV/atom, obtained with GGA and GGA/vdW, respectively. We also optimized, within GGA/vdW level, the structures corresponding to the cases with two and infinite layers (molecular crystal). The optimized geometries for these cases remain unchanged when contrasted with the single-layer one, suggesting that the layers do not strongly interact. The interlayer distance is 3.5 \r{A}.
 
\begin{figure}[!htb]
    \centering
    \includegraphics[scale=0.6]{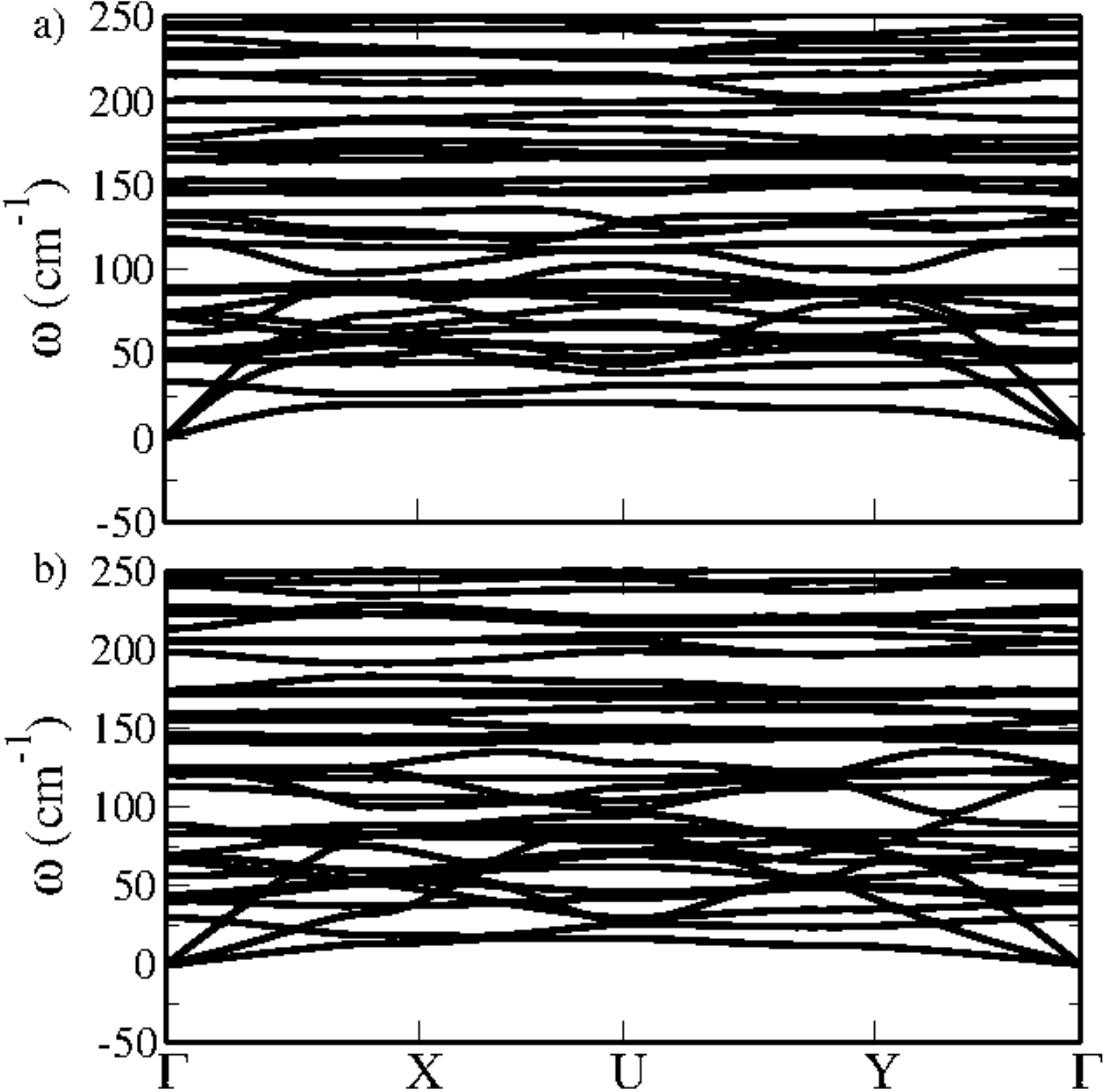}
    \caption{Phonon dispersion band structure along the high symmetric k-point path in the first Brillouin zone for a) GGA and b) GGA/vdW approaches. The high symmetry k-points are: $\Gamma$(0.0, 0.0, 0.0), $X$(0.5, 0.0, 0.0), U(0.5, 0.5, 0.0), Y(0.0, 0.5, 0.0), and $\Gamma$(0.0, 0.0, 0.0).}
    \label{fig:disper}
\end{figure}

To investigate the 2DP structural stability, we performed phonon dispersion and AIMD calculations. For the phonon dispersion, we considered the following high symmetry k points: $\Gamma \rightarrow X \rightarrow U \rightarrow Y \rightarrow \Gamma$. Figure \ref{fig:disper} shows the phonon dispersion spectra for \ref{fig:disper}(a) GGA and \ref{fig:disper}(b) GGA/vdW approximations. In this figure, we notice the absence of negative frequencies, which suggests that the 2DP is stable at T=${}0$ K.

One can also realize that the spectra slightly differ between the two approaches. The acoustic modes close to the $\Gamma$ point are better described in the GGA/vdW approach since one of the three modes is quadratic, as expected for 2D systems \cite{diboron}. The first optical mode for GGA and GGA/vdW occurs near 27 cm$^{-1}$. The three modes are almost horizontal at the $\Gamma$ point, resulting in lower group velocity. This trend is also expected for 2D systems, in which the acoustic modes are more important for the thermal transport process \cite{Balandin2012}.   

\begin{figure}[!htb]
    \centering
    \includegraphics[scale=0.25]{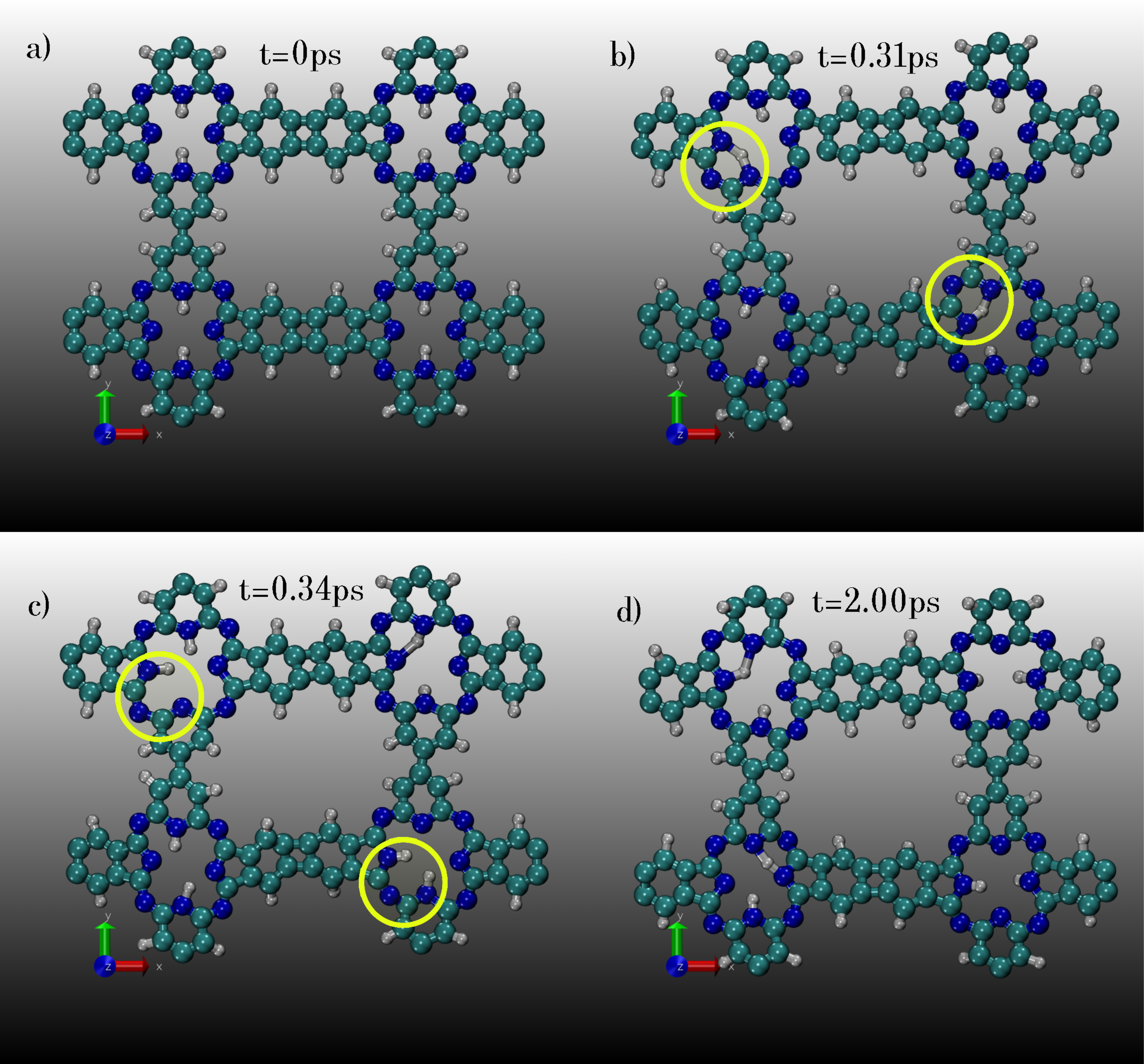}
    \caption{GGA representative AIMD snapshots for the thermal stability simulation of 2DP at 2000 K. The simulation time increases from a) to c) panels. Each panel shows the 2DP lattice configuration atÇ a) 0.00 ps, b) 0.33 ps, c) 2.00 ps, and d) 2DP side view at 2.00 ps).}
    \label{fig:md}
\end{figure}

The temperature effects on the 2DP stability were studied through AIMD simulations. Here, we used an NVT ensemble for temperatures up to 2000 K. Figure \ref{fig:md} show representative AIMD snapshots for GGA results and a total time of 2 ps. For this AIMD simulation, the 2DP unit cell was replicated 2 times along the x and y-directions. We can notice that in the last snapshot (at 2 ps of simulation), the 2DP structure remains similar to the initial one, but presents some buckling caused by the thermal effects. Consequently, we can argue that the 2DP remains stable in high-temperature regimes. 

Another relevant result observed in the AIMD simulations is the proton transfer (site-specific hydrogen exchanges). It occurs when one hydrogen atom, initially bonded to a nitrogen atom, gains sufficient energy to break the bond and form a new bond to next-neighboring nitrogen. This effect was observed from AIMD simulations when two hydrogen atoms are transferred to their next-neighboring nitrogen site at 0.33 ps, as indicated by the yellow circles in Figures \ref{fig:md}(b) and \ref{fig:md}(c). The initial 2DP atomic configuration in Figure \ref{fig:md}(a) differs from the final one in Figure \ref{fig:md}(c), characterizing the proton transfer process. Importantly, this process was extensively investigated in the literature for molecular systems. To the best of our knowledge, this is the first report on the site-specific hydrogen exchange process in 2D crystals.

\subsection{2DP Electronic and Optical Properties}

\begin{figure}[!htb]
    \centering
    \includegraphics[scale=0.4]{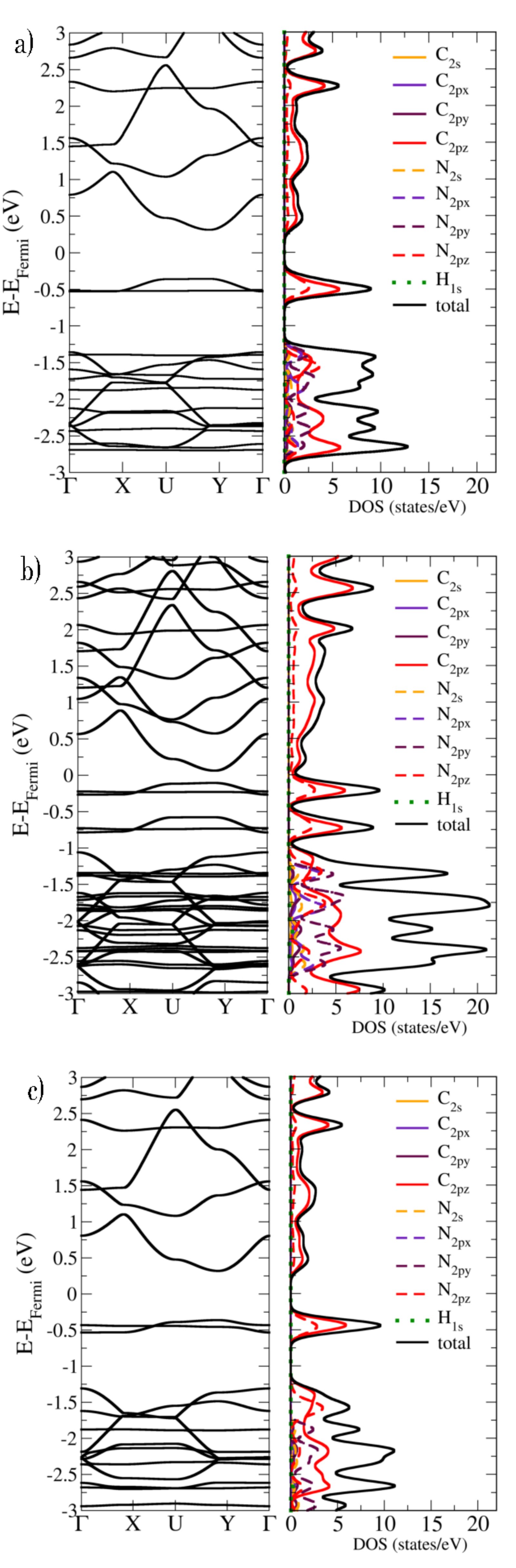}
    \caption{GGA/vdw electronic band structure and the corresponding Projected Density of States (PDOS) for the 2DP systems in (a) single-layer, (b) bi-layer, and (c) bulk configurations.}
    \label{fig:bands_vdw}
\end{figure}

Now, we discuss the electronic and optical properties of the 2DP systems studied here at the GGA/vdW level. Figure \ref{fig:bands_vdw} shows the band structure configuration and the corresponding projected density of states (PDOS) for: \ref{fig:bands_vdw}(a) single-layer, \ref{fig:bands_vdw}(b) bi-layer, and \ref{fig:bands_vdw}(c) bulk (molecular crystal) cases. It is worthwhile to stress that GGA and GGA/vdW band structures are very similar. For this reason, hereafter, we present just the results obtained using the GGA/vdW approximation. 

For the single-layer case (see Figure \ref{fig:bands_vdw}(a)), 2DP has a direct electronic bandgap of 0.65 eV, close to the $Y$-point. Moreover, we notice that for the highest occupied crystalline orbital (HOCO), there is a band splitting (0.2 eV) between the $X$ and $Y$ k-points. Besides HOCO, the lowest unoccupied crystalline orbital (LUCO) and other conduction states are predominantly formed by 2p$_z$ atomic orbitals of the carbon and nitrogen atoms. In contrast, the states below HOCO are formed by contributions from all atomic orbitals.  

For the bi-layer case, we can see from Figure \ref{fig:bands_vdw}(b) that HOCO and HOCO-1 are identical due to the additional layer. The interaction between the layers induces a shift in the valence and conduction states. In this system, the electronic band gap decreases to 0.15 eV, and the corresponding PDOS significantly changes. For the bulk system, we observe that the band structure, and the corresponding PDOS, are very similar to the single-layer case with the electronic bandgap of the bulk phase of 0.66 eV. The same pattern was observed for the LUCO.

\begin{figure}[!htb]
    \centering
    \includegraphics[width=0.7\linewidth]{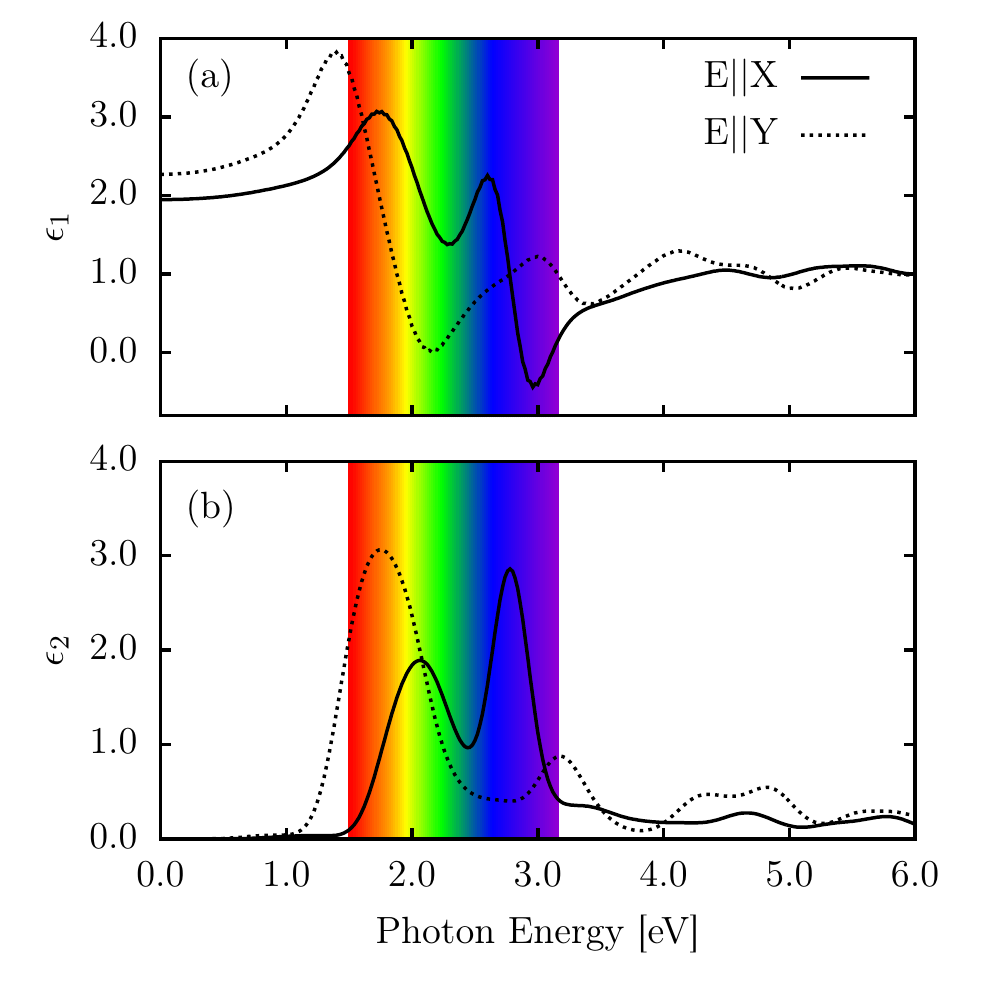}
    \caption{Real ($\epsilon_1$) and imaginary ($\epsilon_2$) parts of the dielectric constant as a function of photon energy for single-layer 2DP. This figure contrasts the results for $\epsilon_1$ and $\epsilon_2$ considering different orientations for the external electric field.}
    \label{fig:epsilon}
\end{figure}

In Figure \ref{fig:epsilon} we present the real ($\epsilon_1$) and imaginary ($\epsilon_2$) parts of the dielectric constant, as a function of photon energy, when subjected to a polarized external electrical field along the x and y-directions. We also investigated the case where the external electrical field is polarized along the z-direction. This interaction produced very small optical coefficient intensities and is not presented in Figure \ref{fig:epsilon}. The real and imaginary components are associated with dispersive and absorptive optical coefficients. Each peak in the spectrum corresponds to an optical activity between allowed transitions of the electronic states, as shown in Figure \ref{fig:bands_vdw}. 

For $\epsilon_1$, one can see in Figure \ref{fig:epsilon} that this value is constant for photon energies between 0 and 1 eV. This trend indicates that 2DP does not present optical activity within the infrared range. For photon energies greater than 4 eV the $\epsilon_1$ value tends to be constant, suggesting a weak optical activity in the violet region. Therefore, the optical activity of single-layer 2DP are more pronounced in the visible region. The X component of the real part of the dielectric function presents negative values for photon energy close to 3 eV. This result indicates that the 2DP layer, when interacting with an external electric field, presents a metallic signature \cite{Alu2006}. 

We can also observe that the dispersion trend in the spectra is isotropic along the x and y-directions. The values for static dielectric constant in each direction can be obtained in the long-wavelength limit (when the photon energy drops to zero) \cite{pentadiamond}. In this sense, $\epsilon_X(0)=2.0$ and $\epsilon_Y(0)=2.3$, for the x and y-directions, respectively. For $\epsilon_2$, we notice in Figure \ref{fig:epsilon} that the optical activity associated with the first interband transitions along the x and y-directions starts for photon energies close to 1.5 and 1.1 eV, respectively. The optical activity occurs with more intensity within the visible region ranging from 1.6 to 3.3 eV.   

\begin{figure}[!htb]
    \centering
    \includegraphics[width=0.7\linewidth]{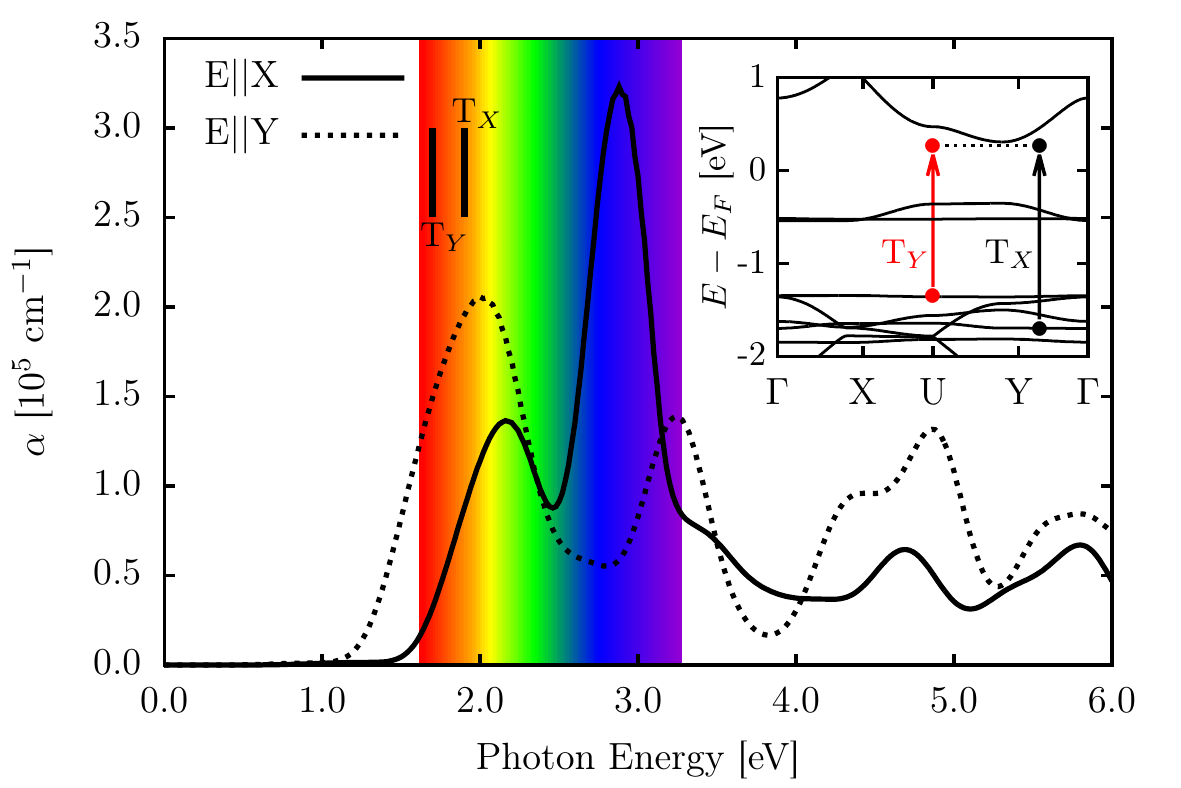}
    \caption{Absorption coefficient $\alpha$ as a function of photon energy for the 2DP single-layer case. The first allowed optical transitions $T_X$ and $T_Y$ are highlighted in black and red colors, respectively. For clarity, the inset panel illustrates the $T_X$ (black) and $T_Y$ (red) transitions in the electronic bad structure.}
    \label{fig:absorption}
\end{figure}

In Figure \ref{fig:absorption} we present the absorption coefficient for 2DP as a function of photon energy for an external electrical field polarized along the x and y-directions. The first allowed optical transitions $T_X$ and $T_Y$ along the x and y-directions, respectively, differ by 0.2 eV. For the sake of clarity, in the inset panel in Figure \ref{fig:absorption} illustrates the $T_X$ (black) and $T_Y$ (red) transitions in the electronic bad structure. These transitions are also highlighted in the absorption coefficient curves as $T_X$ (black bar) and $T_Y$ (red bar) for photon energies of 1.7 and 1.9 eV, respectively. In the two directions of the applied electric field, we notice a moderate optical activity within the visible range.

We can explain the difference between the electronic and optical gaps based on the PDOS results since some electronic transitions are not optical allowed due to symmetry constraints. The first optical transitions occur for the crystalline orbitals below the HOCO, as shown in Figure \ref{fig:absorption} for $T_X$ and $T_Y$ cases. 

\begin{figure}[!htb]
    \centering
    \includegraphics[width=0.7\linewidth]{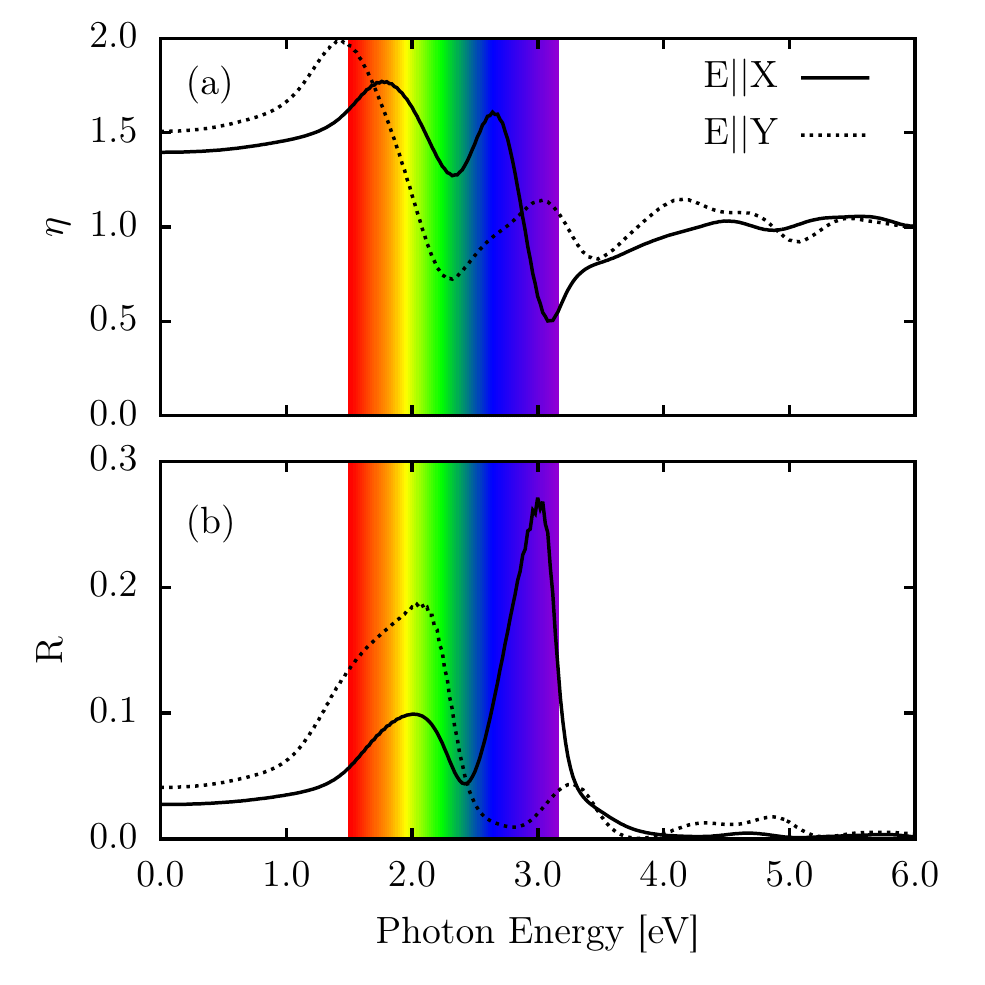}
    \caption{Refractive index ($\eta$) and reflectivity ($R$) as a function of photon energy for 2DP single-layer.}
    \label{fig:refle}
\end{figure}

Now, we discuss the refractive index ($\eta$) and reflectivity ($R$) for the 2DP as a function of photon energy, as shown in Figure \ref{fig:refle}. The $\eta$ index along the x and y-directions is almost constant within the infrared and ultraviolet ranges. We can expect maximum light absorption for photon energies corresponding to the visible region. The maximum refraction occurs near the infrared-visible edge. $R$ also shows more intense activity within the visible region, but the maximum intensity is smaller than 0.3. This value indicates that only 30$\%$ of the incident light will be reflected when interacting with 2DP. This feature suggests that 2DP might be useful in photovoltaic applications.  

We have also analyzed the optical properties for the double-layer and bulk 2DP phases, and the results are similar to the single-layer ones. Although the electronic bandgap changes in the double-layer case, the optical gap remains unaffected since no new states are present to produce new allowed optical transitions. The difference that we can point out is that, for the single-layer case, the intensity is higher concerning the double-layer and bulk phases because the number of allowed states increases with the number of layers, which favors several less intense electronic transitions to occur.   

\section{Lithium Adsorption on 2DP}

To investigate the interaction between Li atoms and 2DP, we carried out AIMD simulations with an NVT ensemble at T${}=300$ K, and using dt${}=1.0$ fs, for a total time of 2.0 ps. The AIMD simulations were performed using the vdW functional and SZ basis set. The vdW functional produces the interaction between species with high precision. The SZ basis set, although smaller than the DZP one, allows obtaining an accurate final configuration with lower computational cost. We consider, as initial configuration, one Li atom at the center of largest pore to mimic a lower concentration case. To simulate a high concentration case,  we added five Li atoms in the 2DP structure (one at the largest pore and the other four at the center of the nitrogen rings).

\begin{figure}[!htb]
    \centering
    \includegraphics[scale=0.34]{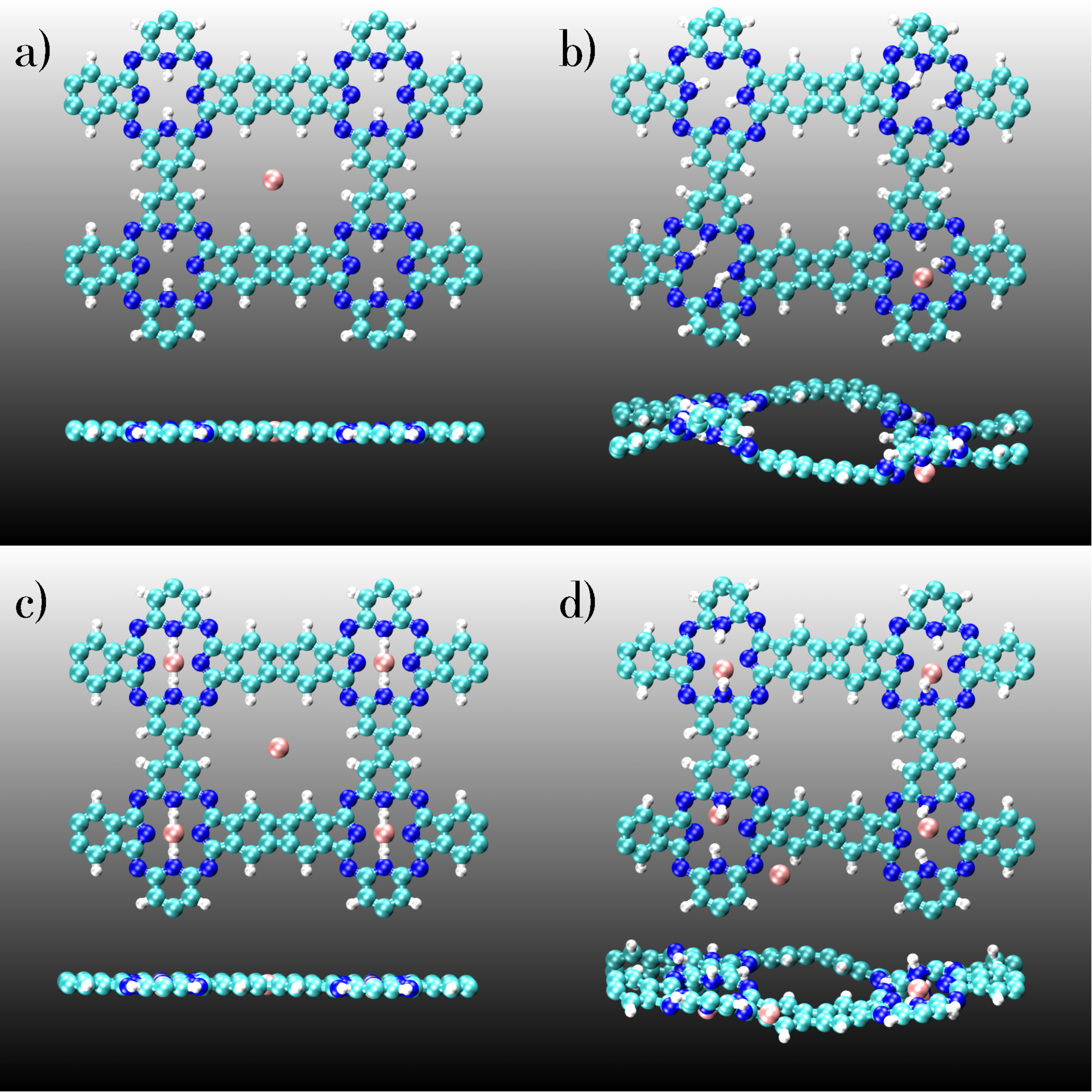}
    \caption{Representative AIMD snapshots for the Li adsorption on 2DP for one (a and b panels) and five (c and d panels) Li atoms at 300 K. (a/b) and (c/d) panels show the initial/final states of the simulations.
    \label{fig:md_Li}}
\end{figure}

In Figure \ref{fig:md_Li} we present AIMD snapshot showing the initial (left, (a) and (c)) and the final (right, (b) and (d))) simulation stages. For the low lithium concentration case, we can see that the final configuration exhibits a pronounced buckling. Also, the lithium atom initially located at the largest pore (Figure \ref{fig:md_Li}-a)) moves to the porphyrin region, as shown in Figure \ref{fig:md_Li}-b). This result agrees with other studies with 2D-based porphyrin systems \cite{por_li,por_Li2}. 

Another interesting observation is regarding the proton transfer. In the case shown in Figure \ref{fig:md}, the proton transfer was observed only at high temperatures, around 2000 K. In the presence of Li atoms, we can see a high proton transfer activity at room temperature, as shown in Figure \ref{fig:md_Li}. As mentioned above, in the proton transfer, one hydrogen atom, initially bonded to a nitrogen atom, gains sufficient energy to break the bond and form a new bond to next-neighboring nitrogen. 

In the case of high lithium concentration, the Li atom is located initially at the largest pore (Figure \ref{fig:md_Li}-c)) moves to the porphyrin region (Figure \ref{fig:md_Li}-d)), as occurred in the case of low Li concentration, while the other Li atoms located initially at porphyrin rings tend to remain in these regions, as can be seen in Figures \ref{fig:md_Li} and \ref{fig:md_Li}-d). Interestingly, at high lithium concentrations, there is no proton transfer. 

The Li atom closest to the hydrogen atom binds to nitrogen inhibiting proton transfer due to the electrostatic interactions mediated by the vdW forces. In this case, the hydrogen does not acquire enough energy to break the initial bond. We provide two movies (see Supplementary Material) corresponding to the AIMD simulations for low and high lithium concentrations. 

\section{Conclusions}

In summary, we proposed a new porphyrin-based 2D nanoporous structure, named 2D Porphyrazine (2DP), which is formed by topological assembling H$_5$C$_{13}$N$_{4}$ porphyrins. We considered single and bi-layer, as well as the molecular crystal (bulk) structures. 

We carried out DFT calculations to investigate 2DP structural and thermal stability. The 2DP phonon spectrum shows no negative frequencies, suggesting its good structural stability at T=${}0$ K. Temperature effects on the 2DP structural stability were studied using \textit{ab initio molecular dynamics} (AIMD) simulations. The results show structural integrity up at 2000 K. 

2DP monolayer has a direct electronic bandgap of 0.65 eV. For the bi-layer case, the electronic band gap decreases to 0.15 eV. For the bulk system, we observed that the band structure, and the corresponding PDOS, are very similar to the single-layer case. The electronic bandgap of the bulk phase is 0.66 eV. 2DP does not present optical activity within the infrared range. Moreover, it has a weak optical activity in the violet region, and most light absorption is in the visible spectrum.

We also performed AIMD simulations to study the interaction between 2DP and Li atoms. We considered, as initial conditions, two cases: one and five Li, to mimic low and high lithium concentration cases.    

For the low lithium concentration case, the lithium atom initially located at the pore with higher diameter moves to the porphyrin region. In the case of high lithium concentration, the Li atom located initially at the largest pore also moves to the porphyrin region, while the other located in the porphyrin region remains close to its initial position. Low Li concentration favors proton transfer (even at low temperature), while high concentration suppresses it.

\input{main.bbl}
\end{document}

%% file: main.bbl
\providecommand{\noopsort}[1]{}\providecommand{\singleletter}[1]{#1}%